\documentclass[conference]{IEEEtran}
\IEEEoverridecommandlockouts

\usepackage{graphicx}
\usepackage{amsmath,amssymb}
\usepackage{booktabs}
\usepackage{tikz}
\usepackage{url}
\usepackage[hidelinks]{hyperref}
\usepackage{cite}

\title{PHP2Shell and the Future of AI-Discovered Vulnerability Disclosure: Evidence, Verification, and Human Authority}

\author{
	\IEEEauthorblockN{
		Toqeer Ali Syed,
		Asadullah Abdullah Khan\textsuperscript{*}
	}
	
	\IEEEauthorblockA{
		Faculty of Computer and Information System,
		Islamic University in Madinah,
		Kingdom of Saudi Arabia\\
		Email: toqeer@iu.edu.sa; Asadullahkhanomarkhil@gmail.com; 471014356@stu.iu.edu.sa\\
		Corresponding author email: Asadullahkhanomarkhil@gmail.com
	}
}

\begin{document}
\maketitle

\begin{abstract}
AI-assisted vulnerability discovery can produce technically plausible findings faster than maintainers can validate, coordinate, and safely disclose them. The resulting challenge is not only to establish whether a vulnerability claim is credible, but also to ensure that its supporting evidence, human authorization, intended recipients, and applicable policy remain bound to the same case revision. We present FedTrust-CVDv2, a federated coordinated vulnerability disclosure framework that separates evidence validation, interorganizational coordination, human governance, and permissioned auditing. The framework binds independently signed observations to exact evidence revisions and scopes external actions to approved payloads, audiences, expiry, policy, and revocation state while preserving existing CVE Numbering Authority (CNA) roles. A local reference implementation accepted all 25 valid fixtures and rejected all 650 invalid fixtures across 27 constructed policy families; six single-safeguard ablations admitted 25--75 invalid fixtures. Supporting experiments exercised signed messaging, authenticated encryption, audit certificates, and finite quorum-set conditions. A PHP2Shell case formulation illustrates how runtime evidence, patch state, emergency coordination, and CNA authority can be represented without attributing unsupported claims to AI or treating audit integrity as vulnerability truth. This short version focuses on the architecture, evidence-to-action contract, selected conformance results, and case formulation. An expanded version containing the full implementation, extended experiments, operational analysis, and additional results will be submitted to a journal.
\end{abstract}


\section{Introduction}
AI-assisted security research changes the economics of vulnerability discovery. A model can generate code references, candidate exploit narratives, and remediation suggestions at machine speed, while maintainers still need to reconstruct the relevant environment, distinguish a defect from an exploitable condition, and decide who may receive sensitive details. OpenAI and Anthropic already describe human review and coordinated-disclosure controls for AI-origin findings \cite{openai_policy,anthropic_policy}. Recent benchmarks likewise show that detection, proof-of-concept generation, reproduction, and patching are distinct capabilities rather than a single notion of ``success'' \cite{secbench,llmrepro,cybergym,bountybench}.

The problem becomes sharper when a finding crosses organizational boundaries. Coordinated vulnerability disclosure (CVD) guidance from CERT/CC, NIST, and FIRST defines reporting and coordination responsibilities \cite{certguide,nist216,first_multiparty}, while Vultron and state-based CVD research formalize multi-party case exchange and workflow state \cite{vultron_blog,vultron_proto,state_model}. Empirical disclosure studies also show that delivery, acknowledgment, acceptance, and remediation must be measured separately \cite{vanhove}. Existing provenance, advisory, and audit technologies provide important building blocks: in-toto and SLSA bind statements to artifacts \cite{intoto,slsa,intoto_attest}; CSAF/VEX, TLP, SSVC, and security.txt support product status, handling expectations, prioritization, and contact discovery \cite{csaf,rfc9116,tlp,ssvc}; Certificate Transparency illustrates append-only accountability \cite{ct}. Permissioned-blockchain disclosure and AI-plus-ledger security frameworks further show that decentralized governance and tamper-evident records are established ideas \cite{amirov,zer0n,syed_icca,syed_energy}.

FedTrust-CVDv2 therefore does not claim novelty for federation, blockchain logging, provenance, or human review individually. Its contribution is a case-level \emph{evidence-to-action authorization contract}: an external action is permitted only when evidence, independent observations, human authority, payload, audience, policy, expiry, revocation, and institutional prerequisites refer to the same current revision. This paper makes three compact contributions: (1) a four-layer architecture for evidence, federation, governance, and audit; (2) a deterministic action contract with revision-bound human authorization; and (3) a selected experimental evaluation plus a PHP2Shell case formulation. The full implementation and broader operational evaluation are reserved for the expanded journal version.

\section{Background and Related Work}
\subsection{Disclosure, validation, and institutional authority}
CVD and CVE publication are related but distinct processes. The CVE Program delegates assignment and record publication to authorized CNAs, and its operational rules and JSON schema define institutional and record-level requirements \cite{cna_rules,cve_schema}. Disclosure timing is a policy decision rather than a universal timer; Project Zero and AI-lab policies illustrate different schedules and escalation conditions \cite{projectzero,anthropic_policy}. FedTrust-CVDv2 consequently treats vendor notification, public advisory release, CVE reservation, and CVE publication as separate events.

Reproduction must also distinguish environment completion from security impact. A crash, timeout, or tool success message does not by itself establish remote code execution or causal attribution. Prior benchmarks and reproduction studies motivate exact-build identity, controlled replay, security-specific oracles, and matched benign or fixed counterfactuals \cite{secbench,llmrepro,cybergym}. The framework carries those artifacts forward into the disclosure decision rather than reducing them to a narrative summary.

\subsection{Provenance, federation, and audit}
Software provenance mechanisms bind claims to artifacts and authorized steps \cite{intoto,slsa,intoto_attest}. FedTrust-CVDv2 reuses that principle for vulnerability evidence, while keeping semantic validity separate from cryptographic integrity. Similarly, decentralized disclosure work demonstrates permissioned governance and authenticated writers \cite{amirov}, and Zer0n combines AI-assisted vulnerability analysis with blockchain-backed report integrity \cite{zer0n}. Related agentic-security work has applied blockchain-backed audit ledgers to autonomous CI/CD defense and, more recently, cryptographically signed agent attestations with permissioned release-policy enforcement for secure software supply chains \cite{syed_icca,syed_energy}. These systems motivate attributable records and enforceable policy checkpoints, while the present design centers the transition from independently reviewed vulnerability evidence to a narrowly authorized disclosure action.

\section{FedTrust-CVDv2 Framework}
\subsection{Architecture and participants}
\begin{figure*}[t]
    \centering
    \includegraphics[width=0.96\textwidth]{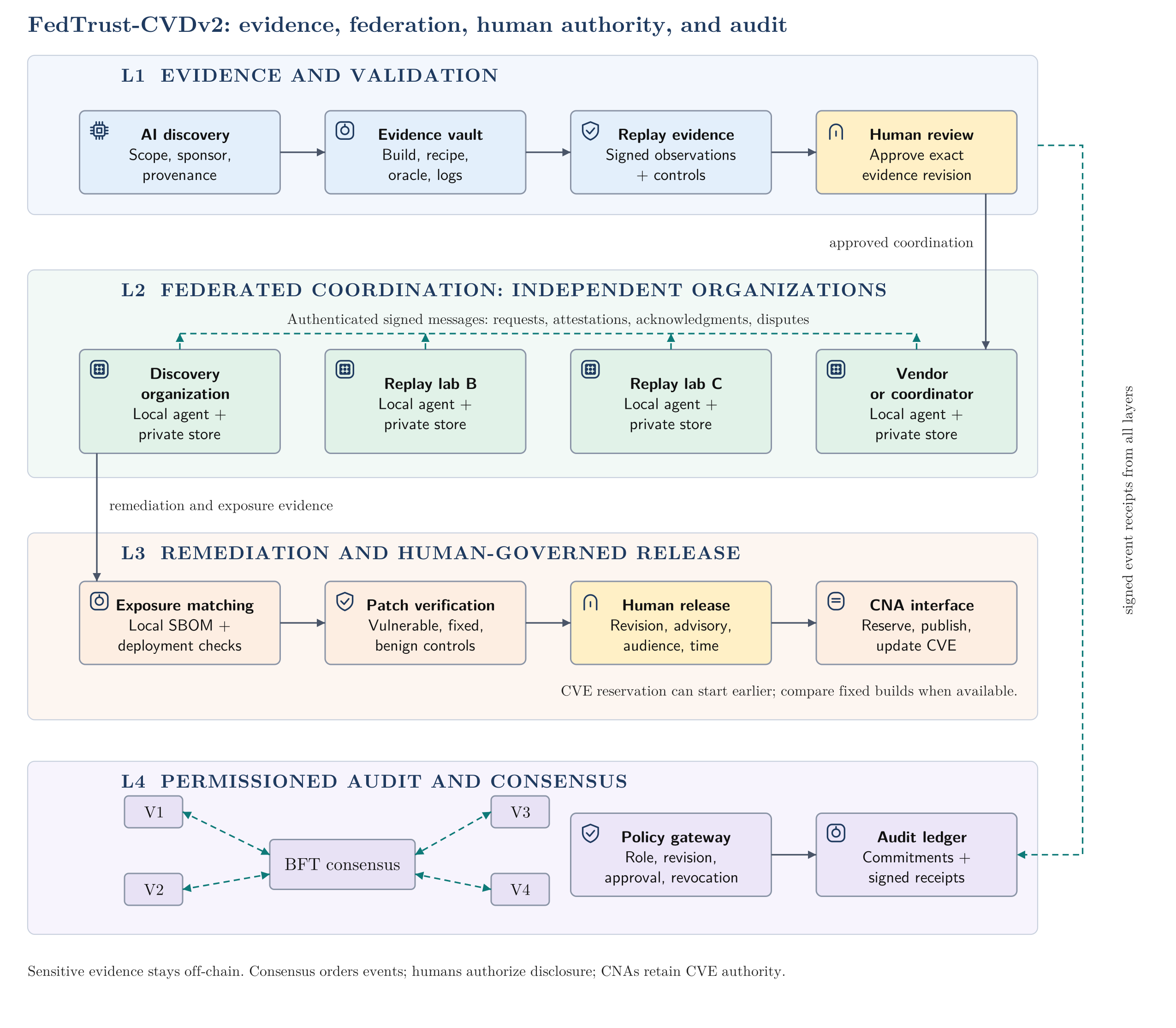}
    \caption{FedTrust-CVDv2 architecture. Evidence, federation, human-governed release, and permissioned audit are separated so that a decision in one layer does not silently grant authority in another. Sensitive evidence remains off-chain.}
    \label{fig:architecture}
\end{figure*}

Figure~\ref{fig:architecture} summarizes four logical layers. The evidence layer records the discovery context, target identity, replay recipe, security oracle, observations, controls, and evidence revision. Independent organizations exchange authenticated, signed coordination messages while retaining local raw artifacts and inventories. Human reviewers authorize consequential actions, and CNAs retain authority for official CVE operations. A permissioned audit plane records accepted workflow events and receipts without requiring sensitive evidence to be placed on-chain.

The main participants are a discovery organization, one or more independent replay organizations, the affected maintainer or vendor, an optional coordinator, downstream operators, and an authorized CNA. Organizational independence is a policy assertion and must not be manufactured by running two agents under one organization. Human and CNA roles are explicit and cannot be satisfied by an AI identity.

\subsection{Evidence-to-action contract}
\begin{figure*}[t]
    \centering
    \includegraphics[width=0.92\textwidth]{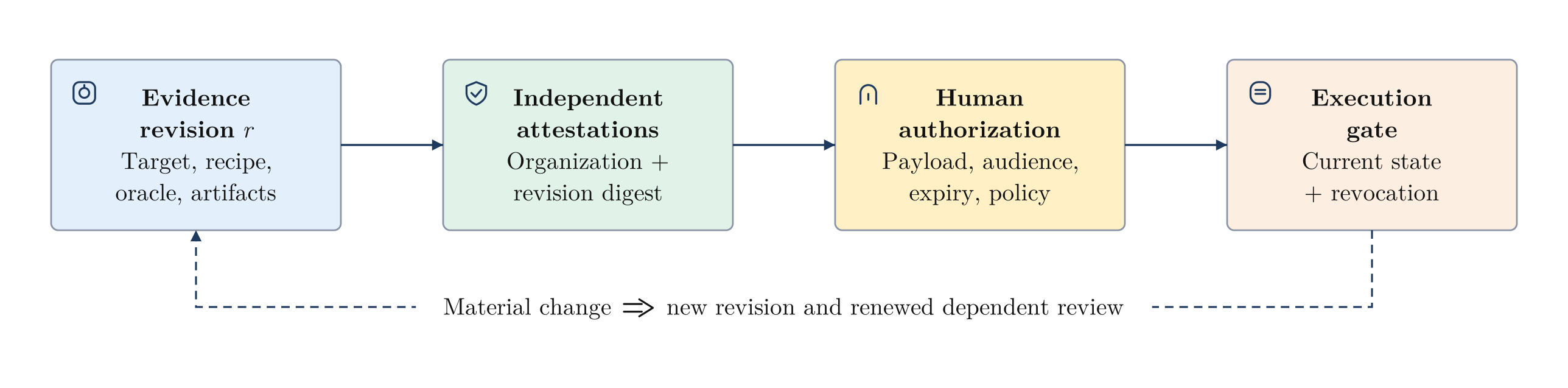}
    \caption{Evidence-to-action chain. Independent attestations and human authorization are bound to an evidence revision; a material change returns the case to renewed validation and review.}
    \label{fig:chain}
\end{figure*}

For case $c$ and revision $r$, let $E_r$ be the evidence capsule and $h_r=H(\mathrm{Canonical}(E_r))$ its digest. A human authorization binds the case and evidence digest to an action, intended audience $U$, approved advisory digest $h_A$, expiry, and policy revision. A material change to the build, oracle, observed impact, evidence digest, recipient set, payload, or policy creates a new revision and invalidates dependent approvals. Figure~\ref{fig:chain} summarizes this dependency: authority does not automatically survive a material evidence change.

An external request is represented as
\begin{equation}
 x=(c,r,h_r,t,U,h_A,p,\tau),
\end{equation}
where $t$ is the operation type, $p$ the policy version, and $\tau$ the requested execution time. The deterministic gate evaluates
\begin{equation}
 \mathrm{Permit}(x,S,R)=I\land E\land V\land H\land B\land F\land C,
\end{equation}
where $I$ covers eligible identities, $E$ scope and evidence freshness, $V$ action-appropriate verifier evidence, $H$ human authority, $B$ exact field binding, $F$ current state and revocation compatibility, and $C$ institutional prerequisites. A valid report hash alone is therefore insufficient: changing the recipient set or advisory payload causes $B$ to fail even when the underlying vulnerability evidence is unchanged.

Independent replay uses distinct organizations and matched controls. For verifier $j$, let $V_j$ denote the vulnerable-build signal, $N_j$ a matched benign-control signal, and $F_j$ a fixed-build signal when a candidate fix exists. The strict differential path uses
\begin{equation}
D_j=V_j\land\neg N_j\land\neg F_j.
\end{equation}
A missing fixed build does not make an initial private report impossible; it instead selects a narrower action policy. Public release can require stronger evidence without forcing every case into a patch-first sequence.

\subsection{Threat model and action policies}
The adversary may fabricate findings, alter evidence, replay an earlier approval, substitute recipients or payloads, register multiple agents under one organization, or exploit stale credentials. Untrusted repository text, logs, and advisory material may also contain prompt-injection content. The framework therefore treats source text as data rather than policy instructions and requires tool use to pass through a deterministic broker. AI services may summarize evidence, suggest recipients, or draft remediation, but cannot sign as a human, enlarge their own network scope, alter an embargo, or invoke a production CVE publication endpoint.

The protected assets are embargoed technical details, verifier infrastructure, private inventories, contact integrity, and decision history. The main security objective is not to make every finding true; it is to prevent a technically or procedurally stale claim from silently acquiring authority. Different actions consequently have different gates. A private vendor report may proceed with credible but incomplete evidence after human review, while a public advisory can require stronger controls, exact payload binding, and multiple designated human signers. Emergency defensive notices are modeled as explicit policy exceptions rather than timer-driven publication. A denial by the strict gate means that a requested action lacks its required authorization state; it does not necessarily mean that the underlying vulnerability is false.

\begin{table}[t]
\caption{Examples of action-specific evidence and authority.}
\label{tab:actions}
\centering
\scriptsize
\setlength{\tabcolsep}{2pt}
\begin{tabular}{p{0.19\columnwidth}p{0.40\columnwidth}p{0.25\columnwidth}}
\toprule
Action & Minimum evidence state & Authority\\
\midrule
Private vendor report & Credible observations, scope checked, uncertainty labeled & Responsible human reviewer\\
Add recipient & Dependency/exposure basis and verified contact & Vendor/coordinator delegate\\
Public advisory & Reviewed payload, current revision, controls or recorded exception & Designated human signers\\
CVE publication & CNA-accepted record and public supporting reference & Authorized CNA workflow\\
Emergency notice & Credible urgent signal; reduced defensive payload & On-call human authority\\
\bottomrule
\end{tabular}
\end{table}

This action separation is important for liveness as well as safety. A newly discovered issue may need prompt private notification before a patch or second independent replay exists; forcing the strict public-release gate onto that early stage would obstruct coordination. Conversely, later public release should not inherit the authority of an earlier private notice. The framework therefore treats patch availability, vendor acknowledgment, CVE reservation, and public release as related but orthogonal case facts rather than one irreversible linear status.

\subsection{Federation and permissioned audit}
Federation messages bind sender, recipient, case alias, revision digest, message type, sequence, nonce, expiry, and payload digest. Raw exploit material and private inventories remain encrypted off-chain. The local design uses authenticated encryption and signed envelopes; production deployments must additionally provide durable replay state, key lifecycle management, and secure transport. AES-GCM and Ed25519 are used as standard primitives \cite{gcm,eddsa}, while canonical JSON should use a pinned cross-language profile such as RFC~8785 \cite{jcs}.

Where shared audit control is needed, an existing BFT ordering substrate can be used rather than inventing a new consensus algorithm. Hyperledger Fabric exposes BFT-capable ordering, and PBFT provides the classical replicated-state-machine basis \cite{fabric,pbft}. For $n=3f+1$ validators and quorum $q=2f+1$, two quorum sets intersect in at least $f+1$ validators. This arithmetic supports certificate reasoning but does not itself demonstrate liveness or deployed Byzantine consensus. Audit finality records that an eligible workflow event was accepted; it does not establish vulnerability truth.

\section{Experimental Method and Selected Results}
The full implementation is reported in the expanded journal version; this short paper retains the experimental design needed to interpret the main result. The local reference artifact uses Python 3.12, Ed25519, SHA-256, and AES-GCM. It models seven application identities and four validator identities and evaluates a deterministic gate over authored policy fixtures. No model API, production inventory, live vulnerability service, or live CVE endpoint is used.

The primary corpus contains 27 policy families with 25 identity-distinct replicas per family, yielding 675 fixtures. One family satisfies the strict gate and 26 introduce a single intended failure, including scope, expiry, dispute, evidence controls, verifier independence, stale revision or digest, human-role failures, audience or payload substitution, revocation, and CNA-role violations. These are constructed policy scenarios rather than independent real-world vulnerabilities.

Six ablations remove exactly one safeguard while retaining all others. Table~\ref{tab:results} gives the central conformance result. The full gate accepted all 25 valid fixtures and rejected all 650 invalid fixtures. Removing organizational independence admitted 50 invalid cases; removing differential controls admitted 75; removing evidence, audience, or payload binding admitted 25 each; and removing revocation checking admitted 50. The ablations show which seeded failure classes depend on each safeguard; they do not estimate operational prevalence.

\begin{table}[t]
\caption{Selected policy-conformance results on the same 675 fixtures.}
\label{tab:results}
\centering
\footnotesize
\begin{tabular}{lrr}
\toprule
Configuration & Valid accepted & Invalid accepted\\
\midrule
Full policy & 25 & 0\\
No independence & 25 & 50\\
No controls & 25 & 75\\
No evidence binding & 25 & 25\\
No audience binding & 25 & 25\\
No payload binding & 25 & 25\\
No revocation & 25 & 50\\
\bottomrule
\end{tabular}
\end{table}

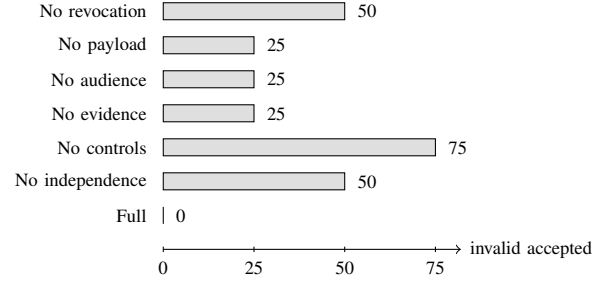
\begin{figure}[t]
\centering
\begin{tikzpicture}[x=0.048cm,y=0.45cm,font=\scriptsize]
  \draw[->] (0,0) -- (82,0) node[right] {invalid accepted};
  \foreach \x in {0,25,50,75} {\draw (\x,0.08)--(\x,-0.08) node[below] {\x};}
  \foreach \name/\val/\y in {Full/0/1,No independence/50/2,No controls/75/3,No evidence/25/4,No audience/25/5,No payload/25/6,No revocation/50/7}{
    \node[anchor=east] at (-2,\y) {\name};
    \draw[fill=gray!25] (0,\y-0.25) rectangle (\val,\y+0.25);
    \node[anchor=west] at (\val+1,\y) {\val};
  }
\end{tikzpicture}
\caption{Invalid fixtures admitted when one safeguard is removed. The denominator is 650 constructed invalid fixtures.}
\label{fig:ablation}
\end{figure}

Figure~\ref{fig:ablation} makes the ablation pattern explicit: controls and organizational independence protect the largest seeded failure families, while evidence-, audience-, and payload-binding each block a distinct substitution class. Revocation checking prevents otherwise valid statements from remaining usable after the relevant identity is revoked. These counts isolate policy dependencies under the authored fixture model and are not population estimates.

Supporting checks exercised 700 signed-message conditions, 400 authenticated-encryption conditions, 120 audit events on four local replicas, and 7,501 quorum-set pairs. The local inbox accepted 100 valid messages and rejected 600 seeded invalid conditions; authenticated encryption completed 100 valid decryptions and rejected 300 altered-key, associated-data, or ciphertext cases. Four local replicas produced identical histories across 120 audit events, and invalid certificate conditions were rejected. Timing on a shared host is intentionally not presented as end-to-end disclosure performance because network transport, persistence, human review, and external services were absent.

\section{PHP2Shell Case Formulation}
The October 2026 PHP security release provides a timely case for separating technical evidence from disclosure authority. The PHP Foundation documented fixes for CVE-2026-109618 and CVE-2026-109619 and identified fixed PHP releases \cite{phpfoundation}. The PHP2Shell research site describes AI-assisted research and exploit development \cite{php2shell}, while AWS separately reported exploitation attempts observed before public disclosure and noted uncertainty about their success on real systems \cite{aws_php}. The evidence therefore supports distinct labels for AI-assisted research, observed exploitation attempts, and confirmed compromise; it does not justify attributing attacker AI use.

\begin{figure*}[t]
    \centering
    \includegraphics[width=0.88\textwidth]{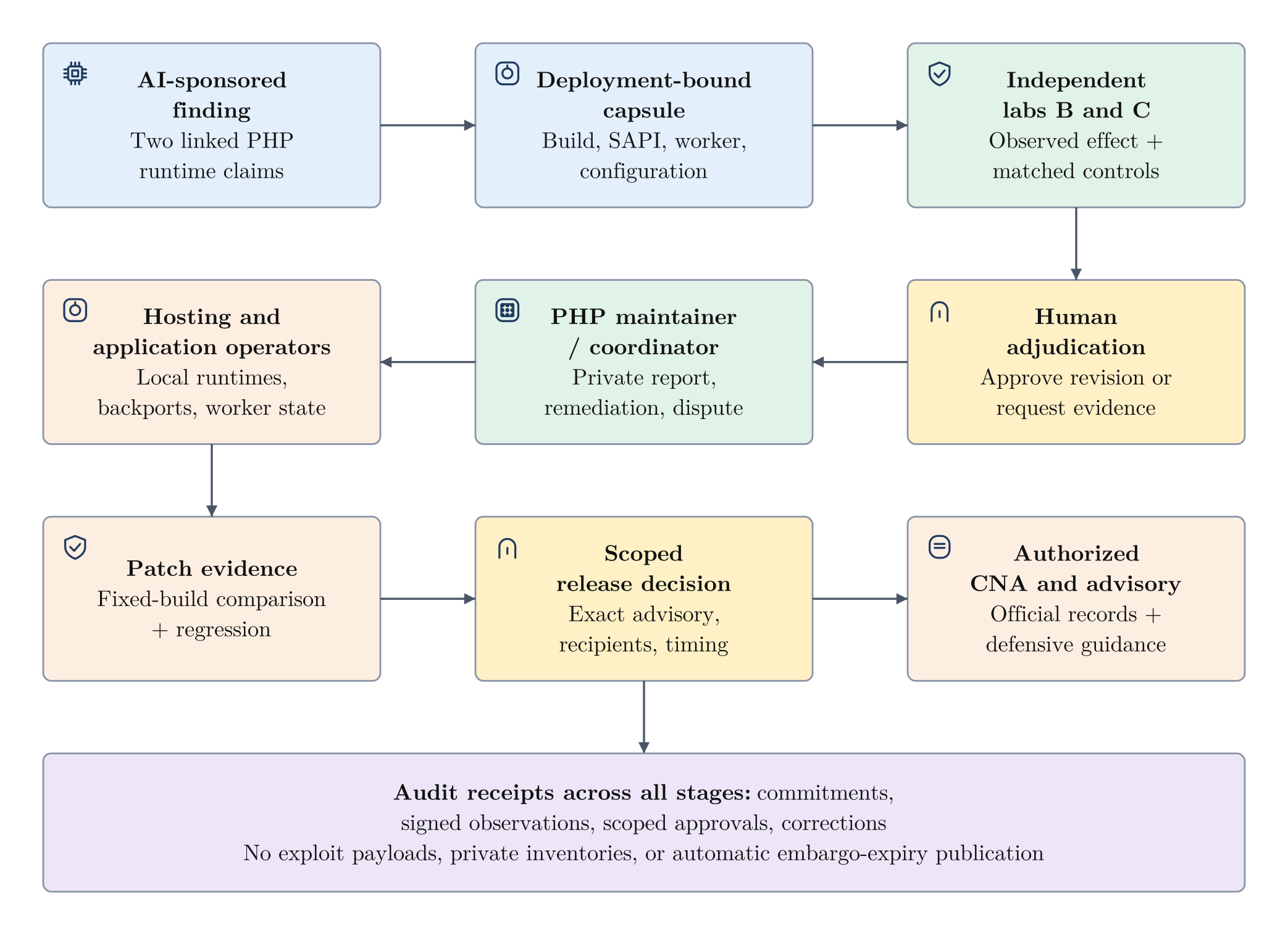}
    \caption{PHP2Shell case formulation under FedTrust-CVDv2. Public facts motivate the runtime findings and fixes; the evidence objects, federation operations, human decisions, and release controls are proposed. Full implementation details and extended case experiments are reserved for the journal version.}
    \label{fig:phpcase}
\end{figure*}

Figure~\ref{fig:phpcase} maps the case into the proposed workflow: an AI-sponsored finding becomes a deployment-bound evidence capsule, independent organizations produce matched observations, a human adjudicates the evidence, and subsequent remediation and release actions remain separately authorized. The diagram is a retrospective formulation of the public facts and proposed control boundaries, not a reconstruction of the maintainers' actual internal process.

The case also shows why a package name or upstream version is not enough. Evidence must identify the PHP build, server API, worker lifecycle, configuration, application image, and distribution or backport state. A mod\_php observation cannot automatically establish the same behavior under PHP-FPM, and installing a patched package does not prove that an old worker process has been replaced. Under FedTrust-CVDv2, independent replay organizations sign observations for the declared deployment, human reviewers authorize a scoped private or public payload, and local inventory nodes return only the minimum routing state permitted by policy. Advisory publication and CVE actions remain distinct and CNA-controlled.

An exploitation signal can raise urgency without changing the underlying evidence revision or granting publication permission. A human may authorize a restricted defensive notice while additional replay continues, and a later public advisory requires its own exact payload and audience authorization. This case formulation is retrospective: the framework was not used by the PHP maintainers, and the study does not claim that FedTrust-CVDv2 would have prevented pre-disclosure exploitation.

\section{Discussion and Limitations}
FedTrust-CVDv2 is best understood as an assurance contract for the transition from evidence to external action. The local evaluation shows that the implemented gate enforces the authored conditions and that removing specific safeguards exposes their seeded failure classes. It does not establish AI vulnerability-detection accuracy, a reduction in real attacks, faster disclosure, improved maintainer trust, or Byzantine safety of a deployed federation.

The design also retains unavoidable human and institutional constraints. Independent organizations can share common-mode errors, reviewers can disagree or become overloaded, private affected-party matching can be incomplete, and recipients can leak information they were legitimately given. CNAs retain assignment and publication authority, and coordinators operate under explicit policies rather than a universal vendor veto or automatic publication timer.

\subsection{Operational evaluation path}
A deployment study should evaluate cases rather than repeated synthetic runs. Relevant outcomes include evidence sufficiency for the requested action, expert-confirmed replay correctness, active human-review effort, notification precision within a consented inventory, and time between first private report, acknowledgment, accepted reproduction, patch availability, advisory publication, and CVE publication. These events should remain separate because successful message delivery is not equivalent to technical acceptance or remediation. Unknown exposure should likewise remain unknown rather than being counted as fixed or unaffected.

A staged pilot should include stale evidence, changed recipient sets, modified release payloads, duplicate or related findings, environment-reconstruction failures, key revocation, malicious instructions embedded in source material, and urgent exploitation reports. Such a study would test whether the evidence contract improves traceability and prevents unsupported external actions under realistic coordination pressure. It would also measure the cost of added review and corroboration rather than assuming that stronger controls are operationally free. These deployment experiments, together with the complete implementation, extended supporting experiments, and additional operational analysis, are reserved for the planned journal version.

\section{Conclusion}
FedTrust-CVDv2 separates AI assistance from disclosure authority by binding technical evidence, independent observations, human decisions, recipients, payloads, policy, and revocation state to an exact case revision. The reference gate rejected all 650 invalid fixtures while preserving all 25 valid fixtures in the constructed corpus, and single-safeguard ablations exposed the intended missing-check failures. The PHP2Shell formulation illustrates how runtime-specific evidence and urgent coordination can be represented without promoting unsupported claims or displacing CNA authority. The next step is operational evaluation under real organizational workflows; that expanded implementation and evaluation will be submitted as a journal article.

\end{document}